\pdfoutput=1

\documentclass[11pt]{article}

\usepackage[]{ACL2023}

\usepackage{times}
\usepackage{latexsym}

\usepackage[T1]{fontenc}

\usepackage[utf8]{inputenc}

\usepackage{microtype}

\usepackage{inconsolata}

\usepackage[utf8]{inputenc} 
\usepackage[T1]{fontenc}    
\usepackage{hyperref}       
\usepackage{url}            
\usepackage{booktabs}       
\usepackage{amsfonts}       
\usepackage{nicefrac}       
\usepackage{microtype}      
\usepackage{xcolor}         

\usepackage{multirow}
\usepackage{float}
\usepackage{pgfplots}
\usepackage{pgfplotstable}

\usepackage{amsmath}
\usepackage{tablefootnote}
\usepackage{algpseudocode}
\usepackage{threeparttable}

\newcommand{\ie}{i.e.,~}

\title{HTEC: Human Transcription Error Correction}

\author{%
  Hanbo Sun \\
  \texttt{sunhanbo@amazon.com} \\
  \And
  Jian Gao \\
  \texttt{gajian@amazon.com} \\
  \And
  Xiaomin Wu \\
  \texttt{wuxiaomi@amazon.com} \\
  \AND
  Anjie Fang \\
  \texttt{njfn@amazon.com} \\
  \And
  Cheng Cao \\
  \texttt{chengcao@amazon.com} \\
  \And
  Zheng Du \\
  \texttt{zhengdu@amazon.com}
}

\pgfplotsset{compat=1.18}

\begin{document}

\maketitle

\begin{abstract}
High-quality human transcription is essential for training and improving Automatic Speech Recognition (ASR) models. Recent study~\cite{libricrowd} has found that every 1\% worse transcription Word Error Rate (WER) increases approximately 2\% ASR WER by using the transcriptions to train ASR models. Transcription errors are inevitable for even highly-trained annotators. However, few studies have explored human transcription correction. Error correction methods for other problems, such as ASR error correction and grammatical error correction, do not perform sufficiently for this problem. Therefore, we propose HTEC for Human Transcription Error Correction. HTEC consists of two stages: Trans-Checker, an error detection model that predicts and masks erroneous words, and Trans-Filler, a sequence-to-sequence generative model that fills masked positions. We propose a holistic list of correction operations, including four novel operations handling deletion errors. We further propose a variant of embeddings that incorporates phoneme information into the input of the transformer. HTEC outperforms other methods by a large margin and surpasses human annotators by 2.2\% to 4.5\% in WER. Finally, we deployed HTEC to assist human annotators and showed HTEC is particularly effective as a co-pilot, which improves transcription quality by 15.1\% without sacrificing transcription velocity.

\end{abstract}

\section{Introduction}
\label{sec:intro}
Automatic Speech Recognition (ASR) models have improved rapidly in recent years, such as wav2vec~\cite{schneider2019wav2vec, baevski2020wav2vec}, Conformer~\cite{gulati2020conformer}, HuBERT~\cite{hsu2021hubert}, and WavLM~\cite{Chen_2022}. These models need a large amount of training data. However, there are only a handful of widely used datasets~\cite{7178964, ardila2020common, TIMIT} of transcribed speech, as it is expensive to obtain large amounts and high-quality transcriptions. To alleviate this problem, many recent works~\cite{baevski2020wav2vec, hsu2021hubert, chung2021w2vbert} combine self-supervised learning with a small volume of transcribed speech, which have made significant improvements in benchmarking datasets, including LibriSpeech~\cite{7178964}. Nevertheless, the data problem is not solved. Large amounts of high-quality transcriptions are crucial to building, improving, and maintaining speech recognition systems, as ASR models often perform much worse in the real world than benchmark results.
  
In practice, human annotators listen to speech and transcribe it into text. These transcriptions are commonly used as the ground truth to build and improve ASR models, and further processed for downstream natural language understanding tasks. However, human annotators can easily make transcription errors due to a lack of domain knowledge, misheard audio, typos, and other factors.

\begin{table*}
  \centering
   \resizebox{0.99\textwidth}{!}{
   
   \begin{threeparttable}
   
    \begin{tabular}{rlll}
    \hline\noalign{\smallskip}
    \multicolumn{1}{l}{Error Type} &
    \multicolumn{1}{l}{Error Causes} &
    \multicolumn{1}{l}{Prevalence} &
    
    \begin{tabular}[c]{@{}l@{}}Annotator Transcripts (\textcolor{red}{red})\\ Gold Transcripts (\textcolor{blue}{blue})\end{tabular} \\
    \noalign{\smallskip}\hline\noalign{\smallskip} 
    
    \multicolumn{1}{l}{Convention Error} &  
    \begin{tabular}[c]{@{}l@{}}Not following pre-defined conventions; \\ Making format error: punctuation, spoken/written \end{tabular} &
    8.57\% &
    \begin{tabular}[c]{@{}l@{}}Order \textcolor{red}{AAA} battery \\
                               Order \textcolor{blue}{triple A.} battery \end{tabular} \\
    \noalign{\smallskip}\hline\noalign{\smallskip}
    
    \multicolumn{1}{l}{Spelling Mistake} & 
    \begin{tabular}[c]{@{}l@{}}Typing too quickly; \\ Not being familiar with the spelling\end{tabular} &
    11.63\% &
    \begin{tabular}[c]{@{}l@{}}What is \textcolor{red}{on} and a half sticks of \textcolor{red}{buttern} \\ 
                                What is \textcolor{blue}{one} and a half sticks of \textcolor{blue}{butter} \end{tabular} \\
    \noalign{\smallskip}\hline\noalign{\smallskip}
    
    \multicolumn{1}{l}{Grammatical Error} &   
    \begin{tabular}[c]{@{}l@{}}Not paying attention to grammar; \\ Not aligning with grammar used in utterance\tnote{*} \end{tabular} &
    11.02\% &
    \begin{tabular}[c]{@{}l@{}}Give \textcolor{red}{my}, um... the latest news \\
                                Give \textcolor{blue}{me}, um... the latest news \end{tabular} \\
    \noalign{\smallskip}\hline\noalign{\smallskip}
    
    \multicolumn{1}{l}{Lack Domain Knowledge} &  
    \begin{tabular}[c]{@{}l@{}}Not familiar with the entities in utterance; \\ Having difficulty understanding complex concepts \end{tabular} &
    18.37\% &
    \begin{tabular}[c]{@{}l@{}}List ingredients for \textcolor{red}{Michelle} cocktail \\
                               List ingredients for \textcolor{blue}{Mitchell} cocktail \end{tabular} \\
    \noalign{\smallskip}\hline\noalign{\smallskip}
    
    \multicolumn{1}{l}{Misheard Audio} & 
    \begin{tabular}[c]{@{}l@{}}Utterance has strong noise, cross-talk, distorted voice; \\ Losing attention to homophone during high-throughput work \end{tabular} &
    50.41\% &
    \begin{tabular}[c]{@{}l@{}}When \textcolor{red}{does} my \textcolor{red}{lost} Amazon order \\
                               When \textcolor{blue}{was} my \textcolor{blue}{last} Amazon order \end{tabular} \\
    
    \noalign{\smallskip}\hline
    
    \end{tabular}
    \begin{tablenotes}\footnotesize
    \item[*] Annotator should transcribe utterance ``as it is'' despite it may contain grammatical mistake, pause filler, or other issues.
    \end{tablenotes}
    \end{threeparttable}
    }
    \caption{Study of Errors Type, Cause, and Prevalence in Human Transcriptions.}
  \label{tab:spokenerrortype}
\end{table*}%

Few studies have explored human transcription error correction. \citet{namazifar2021correcting} have mainly focused on error data augmentation. \citet{Hazen2006AutomaticAA} have proposed a method to correct speech-transcription alignment errors. In other relevant areas, various approaches have been proposed for Grammatical Error Correction (GEC)~\cite{yasunaga2021lm, omelianchuk2020gector} and ASR Error Correction~\cite{namazifar2021correcting, leng2021fastcorrect}. In addition, generic approaches such as fine-tuned sequence-to-sequence models~\cite{lewis2019bart, 2020t5} and in-context learning of LLM~\cite{brown2020language} can be applied. However, these methods perform insufficiently in improving speech transcription as they are designed for written language or ASR, and few-shot learning is not sufficient for such a complicated problem. Firstly, spoken language is spontaneous, fleeting, and casual. As presented in Table~\ref{tab:spokenerrortype}, we collected 1K erroneous transcriptions, and categorized the types and causes of error. The prevalent errors include a lack of domain knowledge and misheard audio. In contrast, annotators make fewer simple errors, such as spelling and grammar errors, which are more common in written language. Secondly, the patterns of human transcription errors are usually sporadic and difficult to predict. Table~\ref{tab:error_dist} presents that the distributions of errors made by humans and ASR models are different. Humans tend to make more insertion errors and fewer deletion errors. 


\begin{table}[!htb]
  \centering
  \resizebox{0.48\textwidth}{!}{
    \begin{tabular}{lccc}
\hline\noalign{\smallskip}
 & Substitution Error & Insertion Error & Deletion Error \\
\hline\noalign{\smallskip}
\begin{tabular}[c]{@{}l@{}}Annotator   \\ Transcription\end{tabular} & 35.0\% & {37.3\%} & {27.7\%} \\
\hline\noalign{\smallskip}
\begin{tabular}[c]{@{}l@{}}ASR \\ Transcription\end{tabular} & 41.7\% & {16.3\%} & {42.1\%} \\
\hline\noalign{\smallskip}
\end{tabular}
}
\caption{Compare Word Error Distribution between Human Transcriptions and ASR Model Transcriptions.}
  \label{tab:error_dist}%
\end{table}%

This paper presents a novel approach called HTEC to improve human transcription quality. HTEC is particularly effective in assisting the human annotator as a co-pilot. We benchmark HTEC with other models and present the results of the deployed product in a user study. HTEC outperforms other baselines by a large margin, with relative WER improvements of 25\%, 12\%, 10\%, and 5\%, respectively, compared to GEC, generic sequence-to-sequence model, LLM in-context learning, and ASR error correction methods. HTEC also outperforms human annotators by 2.2\% to 4.5\%. As a co-pilot, HTEC assists professional annotators to improve transcription quality by a relative 15.1\% of WER. To the best of our knowledge, this paper is one of the first studies in human transcription error correction for spoken language. In summary, the contributions of this paper are:
\\


\begin{itemize}
    \item We formulate the problem of human transcription error correction and categorize the error type and cause.
    \item We propose HTEC, a general two-stage framework with components: Trans-Checker to predict erroneous words and Trans-Filler to fill in positions predicted as errors.
    \item We propose a holistic list of editing actions, including four simple yet useful operations for deletion errors.
    \item We propose a variant of the embedding layer to incorporate phoneme information.
    \item We run extensive experiments and user studies and show HTEC reduces WER by 2.2\% to 4.5\% over human annotators and by 15.1\% when it collaborates with humans as a co-pilot. We will release code to enable researchers to improve ASR models by improving transcription data quality.
\end{itemize}

\section{Related work}
\label{sec:relatedwork}
Recently, multiple error correction models have been developed to refine grammar and spelling errors, such as LM-Critic \cite{yasunaga2021lm} and GECToR \cite{omelianchuk2020gector}. They have achieved top performance in the benchmarking dataset BEA-2019 \cite{bryant2019bea}. LM-Critic introduces a critic method that uses trained language models and a designed Break-It-Fix-It (BIFI) framework \cite{yasunaga2021break}. GECToR adopts a transformer encoder to build a sequence tagger that converts text correction to a text tagging task. The tags include both required editing actions and word candidates related to the actions. In addition, NeuSpell \cite{DBLP:journals/corr/abs-2010-11085} has adapted 10 error correction models by adding contextual representations and synthetic misspelling data.

For ASR error correction, it is popular to modify erroneous words or select replacements from ASR output sequences to reduce WER. ASR systems commonly apply a two-pass paradigm~\cite{rescorer}, where the first pass generates the n-best hypotheses and the second pass re-ranks them using a re-scoring model. Namazifar et al.~\cite{namazifar2021correcting} have studied different types of errors and used masked language models to correct ASR errors. Yang et al.~\cite{DBLP:journals/corr/abs-2208-04641} proposed an ASR error correction method that utilizes the predictions of correction operations. FastCorrect \cite{leng2021fastcorrect} refines the output sentences of ASR systems. It uses a length predictor to guide the correction.

Additionally, it is becoming more popular to utilize LLM in-context learning for various tasks. The commonly used LLMs include GPT-series models~\cite{ouyang2022training, openai2023gpt4}, PaLM-series models~\cite{chowdhery2022palm, anil2023palm}, and open source LLMs such as LLaMA~\cite{touvron2023llama}, Alpaca~\cite{alpaca}, and BLOOM~\cite{workshop2023bloom}. We will also evaluate LLM for comparison.

While grammatical error correction focuses on errors in written language, and ASR error correction improves the output of ASR models, HTEC targets improving human transcription of speech. To our best knowledge, HTEC is one of the first solutions for correcting human transcription errors that have been proven useful in real-world applications as either an autocorrection model in post-processing or a co-pilot tool in assisting human annotators during transcription.

\section{HTEC}
\label{sec:method}
\begin{figure}[!htbp]
\centering
\includegraphics[angle=0,width=0.48\textwidth,keepaspectratio]{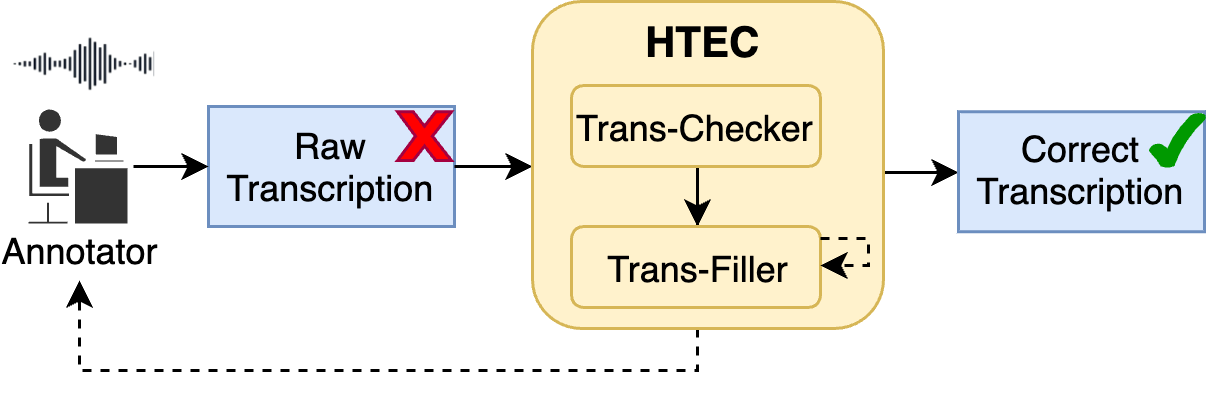}
\caption{Overall framework of HTEC: Trans-Checker detects errors, and Trans-Filler fills the positions where Trans-Checker predicted errors. Trans-Filler may take multiple steps to fill these positions. The annotator can then further edit based on recommendations.}
\label{fig:overview}       
\end{figure}

Given a raw transcription by a human annotator, the goal of HTEC is to generate an accurate transcription that is highly congruent with the gold transcription. The gold transcription is obtained by multiple blind-pass annotators and an additional adjudicator if the annotators cannot reach a consensus. HTEC is a two-stage framework with two main components: Trans-Checker and Trans-Filler. In the first stage, we propose Trans-Checker, a transformer-based error detection model that predicts erroneous words. In the second stage, we propose Trans-Filler, a generative sequence-to-sequence model that fills in the positions of predicted errors by Trans-Checker. HTEC follows the way that human SMEs identify and fix errors in transcription. Figure \ref{fig:overview} shows the overall workflow of HTEC. Besides automatic error correction, HTEC is particularly suitable for assisting human annotators in a human-in-the-loop (HIL) manner. 

\subsection{Trans-Checker}
\begin{figure*}[t]
\centering
\includegraphics[angle=0,width=0.9\textwidth,keepaspectratio]{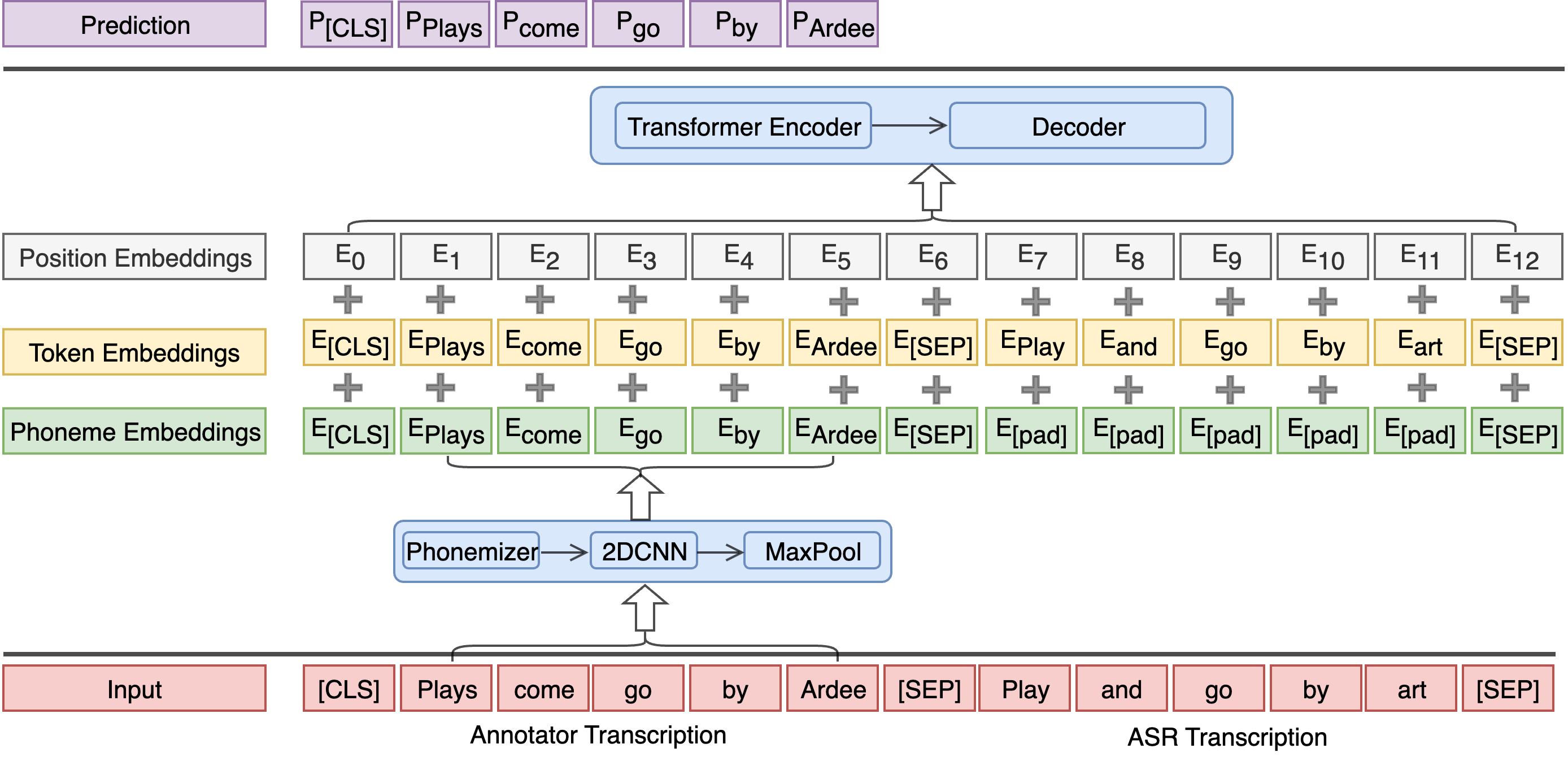}
\caption{Trans-Checker input representation and model structure: The input embeddings are the sum of token embeddings, position embeddings, and phoneme embeddings that are obtained from a CNN and max-pooling layer.}
\label{fig:errdetector}       
\end{figure*}

Trans-Checker predicts the type of word error for each word in a given annotator transcription. The model structure is shown in Figure \ref{fig:errdetector}. The annotator transcription and ASR transcription are concatenated to form the token embedding and position embedding. The phonomizer converts the annotator's transcription to generate phoneme embeddings through a convolutional neural network (CNN) and pooling. The phoneme embeddings are updated with the training of Trans-Checker and CNN parameters. The phoneme embeddings learn the ambiguity between homophones or similar sounding words from training data. The three embeddings are input to the transformer encoder, followed by the decoder, whose output length is equal to the length of the annotator transcription.

\subsubsection{Novel Edit Operations}
The label of each word is the edit operation to correct the word and its surrounding words, if available. The labels are derived by aligning the pair of annotator transcription and gold transcription by minimizing the edit distance~\cite{bryant2017errant}, and then comparing each word in annotator transcription with its counterpart in gold transcription. The widely used edit operations are insertion ($I$), deletion ($D$), and substitution ($S$). Keep ($K$) indicates no error. We propose four correction actions: {$KL$, $KR$, $SL$, $SR$} to replace operation $I$, where $KL$ suggests keeping the word and inserting word(s) to the left of the word, $SR$ suggests substituting the word with a different word and inserting words to the right of the word. Table~\ref{tab:errortypes} presents the operations in detail. The words with any of the four labels are called \textit{anchor word}. These four new labels are simple yet useful and convert the error detection problem to token classification.

\begin{table}[!htbp]
  \centering
  
  \resizebox{0.48\textwidth}{!}{
    \begin{tabular}{ll}
    \hline\noalign{\smallskip}
    Code & Describe Operation to Correct \\
          \noalign{\smallskip}\hline\noalign{\smallskip}
    K     & \textbf{K}eep the word \\
    D     & \textbf{D}elete the word \\
    S   & \textbf{S}ubstitute the word with a different word \\
    \textcolor{blue}{KL} & \textbf{K}eep the word and insert word(s) to the \textbf{L}eft \\
    \textcolor{blue}{KR}& \textbf{K}eep the word and insert word(s) to the \textbf{R}ight \\
    \textcolor{blue}{SL} & \textbf{S}ubstitute the word and insert word(s) to the \textbf{L}eft \\
    \textcolor{blue}{SR} & \textbf{S}ubstitute the word and insert word(s) to the \textbf{R}ight \\
    \noalign{\smallskip}\hline
    \end{tabular}%
    }
    \caption{Proposed Editing Actions to Correct Errors.}
  \label{tab:errortypes}%
\end{table}%

\subsubsection{Phoneme Embedding}
\label{ssub:phoemb} 
As presented in Table~\ref{tab:errortypes}, 50.4\% of transcription errors are related to misheard audio due to low quality of speech or lack of attention from the annotator. We introduce a variant of the embedding layer that incorporates phoneme embeddings as part of the transformer input, which augments the model's ability to correct errors caused by homophonic or similar-sounding ambiguity. Phoneme embeddings are generated by phonemizing annotator transcription into phonemes at the word level. One word in the transcription can be phonemized into multiple phonemes. There are a total of 44 unique phonemes generated by the festival phonemizer \cite{Bernard2021} in English. The maximum number of phonemes per word is set to 20, and average pooling is applied to obtain phoneme embedding for each word. The maximum number of words in a sentence is set to 64. A phoneme pad is appended to each sentence and initialized randomly in model training. Phonemes are converted into phoneme embeddings through a CNN model~\cite{fang2020using} and max pooling layer. The phoneme embeddings are added to positional embedding and input token embedding. The phoneme embedding and CNN parameters are updated along with the encoder and decoder in the main model.

\subsection{Trans-Filler}
\label{sec:method_filler}

\begin{figure}[!htbp]
\centering
\includegraphics[angle=0,width=0.48\textwidth,keepaspectratio]{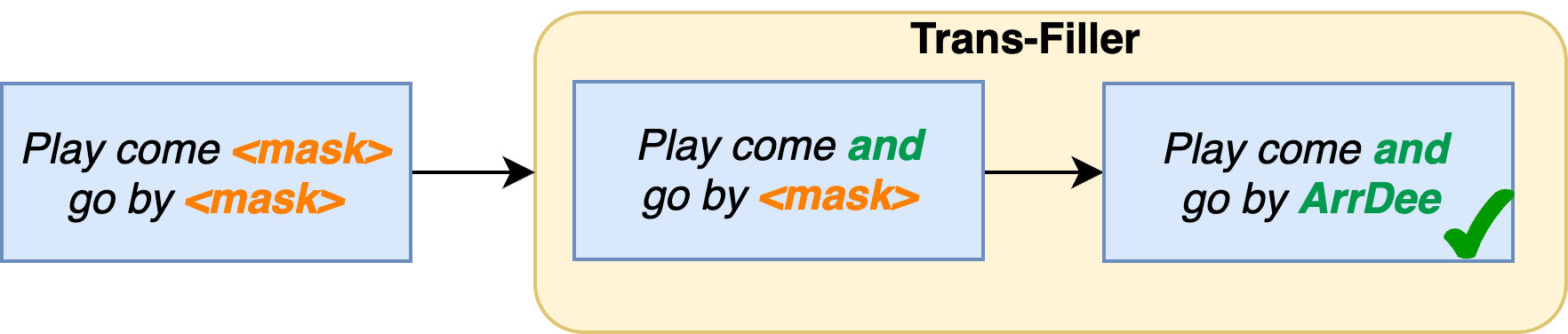}
\caption{The iterative steps of Trans-Filler: It fills the positions that Trans-Checker predicts as errors or that the annotator needs assistance with. These positions are masked by special tokens: \texttt{<masks>}. Trans-Filler fills one mask per iteration until all masks are filled.}
\label{fig:filler}       
\end{figure}

Trans-Filler fills positions where Trans-Checker predicted errors (\ie auto-correction), or annotator is not confident to transcribe (\ie co-pilot correction). It is developed on a sequence-to-sequence generative model as the backbone. The backbone model can be replaced for alternatives such as BART~\cite{lewis2019bart} or T5 models~\cite{2020t5}. We propose an iterative process, as illustrated in Figure~\ref{fig:filler}. One mask is filled in each iteration until all masks are filled. One mask can be filled with one or more words. The decoder can be autoregressive (AR) or non-autoregressive (NAR). In the AR decoder setting, the encoder is trained with one position of right-shifted labels, which are fed into the decoder. AR cannot model distributions whose next-symbol probability is hard to compute, which is more commonly seen in spoken language that is more casual. The AR decoder tends to fill multiple tokens at one position so that more existing errors can be fixed. The NAR decoder tends to fill one token in one position so it introduces fewer new errors. Similar to Trans-Checker shown in figure~\ref{fig:errdetector}, Trans-Filler also takes annotator transcription and ASR transcription as input text.

\subsection{Transformer Architecture}
\label{app: mdoel_arch}
The neural network models implemented in this paper are based on the self-attention Transformer architecture~\cite{NIPS2017_3f5ee243}. Formally, given a sequence of source input tokens that are encoded by one-hot encoding, \ie vector $S = (s_1,\dots, s_I )$ where $s_i \in \textrm{VS}$, the goal is to predict a sequence of target output tokens $T = (t_1,\dots, t_o )$ where $t_o \in \textrm{VO}$. \textrm{VS} and \textrm{VO} are input and output token vocabulary, respectively. 
For Trans-Checker, the token classification task specifically, $S = [f_1|f_2]$ where $f_1$ is the sequence of annotator transcription with length $I_1$, and $f_2$ is ASR transcription  with length $I_2$. $I_1+I_2=I$. Output sequence length is $I_1$, \ie, $T = (t_1,\dots, t_{I_1})$. In contrast, Trans-Filler is a sequence-to-sequence model with an input sequence: 

\scalebox{0.9}{
$S = (s_1,\dots,<mask>,\dots,<mask>,\dots,s_{I_1} )$
}
where $s_i \in \textrm{VS}$. $S$ contains $k$ \texttt{<mask>} to be filled in k iterations. The output sequence is:
\begin{equation}
O = (o_1,\dots,o_J )
\end{equation}
where $J\geq I_1-k$.

Both Trans-Checker and Trans-Filler have two main components: the encoder and the decoder. The encoder transforms the source sequence into a sequence of hidden states by mapping each individual token into a continuous embedding space, adding positional embeddings and  phoneme embeddings. Then processing it through a sequence of self-attention and feed-forward layers:
\begin{equation}
X_{1,\dots,I} = E_{s_{1,\dots,I}} + P_{1,\dots,I} + PH_{s_{1,\dots,I}}
\end{equation}

\begin{equation}
H_{1,\dots,I} = f_{enc}(X_{1,\dots,I})
\end{equation}

where $x_i \in \textrm{V}$, $E$ is the embedding matrix for vocabulary \textrm{VS} and $P_{1,\dots,I}$ is the sequence of positional embeddings. $PH_{s_{1,\dots,I}}$ is the phoneme embeddings matrix for vocabulary $\textrm{VS}$. $f_{enc}(\cdot)$ is the encoder that converts embedding into hidden states.

The decoder defines the distribution of probabilities $\textrm{P}^{'}$ over all items in the vocabulary at each time step $t$.

\begin{align} 
y^{'}_{1,\dots,t-1} &= E_{t_{1,\dots,t-1}} + P_{1,\dots,t-1} + PH_{s_{1,\dots,t-1}} \\ 
H_{t}^{dec} &= f_{dec}(y^{'}_{1,\dots,t-1}, H_{1,\dots,I})E^{T} \\ 
P_{t}^{'} &= softmax(H_{t}^{dec}E^{T}) 
\end{align}
where $f_{dec}(\cdot)$ is the decoder. $t=(1,\dots,I_1)$ for Trans-Checker, and $t \in \textrm{K}$ for Trans-Filler. \textrm{K} is the set of indexes for \texttt{<mask>}.
We optimize the cross-entropy loss across time steps.
\begin{equation}
L_{CE}(P^{'}) = -\sum_{t}log(P^{'}_{t}))
\end{equation}

\section{Experiment}
\label{sec:experiments}
\subsection{Experiment Setting}

\paragraph{Data}
Trans-Checker, Trans-Filler, and HTEC are evaluated on two datasets, a de-identified commercial voice assistant dataset, Conversational AI Agent (CAIA), and a public dataset, MASSIVE~\cite{fitzgerald2022massive}. CAIA contains 383k utterances (490 hours). MASSIVE contains 19k utterances (5 hours). Both datasets are in English and contain annotator standard transcriptions and gold transcriptions. In each dataset, we randomly selected 10\% data as the test set.

\paragraph{Model Architecture} Trans-Checker is a transformer-based token classification model that contains $12$ stacked transformer blocks following the decoder. Phoneme embeddings are added to the input embedding layer. Phoneme embeddings are generated by CNN with a filter size of 3x3 and max-pooling layers. Trans-Filler is a sequence-to-sequence network with $24$ stacked transformer blocks: $12$ for the encoder and $12$ for the decoder. 

\paragraph{Training} To train Trans-Checker, we first generate word-level error labels from pairs of annotator and gold transcriptions using the linguistically-enhanced Damerau-Levenshtein alignment algorithm~\cite{DamerauEdit}. Annotator transcriptions concatenated with ASR text are fed into the model as training inputs. To train Trans-Filler, we compare two model variants: BART with an autoregressive or non-autoregressive decoder. The former can fill multiple words at one position, while the latter tends to perform one-to-one mask filling. During the training of Trans-Checker and Trans-Filler, $10\%$ of the training data is used for validation. The training batch size is set to $64$ with the learning rate scheduler. Models are trained for $30$ epochs with early stopping. 
 
\paragraph{Evaluation}
In addition to human annotators as baseline, we compare HTEC with six SOTA methods in four categories: (1) GECToR and LM-Critic for grammatical and spelling mistake correction; (2) ConstDecoder and Rescorer for ASR error correction; (3) BART as a generic sequence-to-sequence model; (4) Alpaca~\cite{alpaca} for LLM in-context learning. We fine-tune the first three types of models to correct errors in annotator transcriptions. For Alpaca, we experiment with one-shot and few-shot prompts.

Trans-Checker is evaluated by Precision, Recall, F1 and AUC. Trans-Filler is evaluated by Precision, Recall, and F1, in addition to WER. Finally, we evaluate HTEC from end to end, \ie cascading Trans-Checker and Trans-Filler. The transcription WER is used to measure model performance. We call a model "automatable" for transcription correction if $WER_m < WER_a$, where $WER_m$ is the WER of corrected transcription by the model against gold transcription. $WER_a$ is the WER of annotator transcription against gold transcription. An automatable model can reduce WER without human attention.

\subsection{Performance Evaluation}

\subsubsection{Trans-Checker Evaluation}
We experiment with four pre-trained language models: BERT~\cite{devlin2019bert}, ELECTRA~\cite{clark2020electra}, BART~\cite{lewis2019bart}, BERT with phoneme embedding (BERT-Pho). The pre-trained models are fine-tuned for error classification tasks. The evaluation results on the CAIA dataset are shown in Table \ref{tab:errorDetectorAlexa}. Models are evaluated on the test set for precision, recall, macro F1, and AUC. Phoneme embedding improves F1 and AUC 2\%-3\% compared to vanilla BERT and achieves comparable results as BART.

\begin{table}[!htbp]
  \centering  

  \resizebox{0.45\textwidth}{!}{
    \begin{tabular}{lrrrr}
    \hline\noalign{\smallskip}
          & Precision &  Recall &   F1 &  AUC \\
          \noalign{\smallskip}\hline\noalign{\smallskip}
    BERT & $x_p$ & $x_r$ &  $x_f$ & $x_a$ \\
    ELECTRA & -0.015 & +0.001 & -0.003& -0.001 \\
    BART  & -0.095 & +0.056 &  +0.030 & +0.021 \\
    BERT+Pho  & -0.036 & +0.040  & +0.031 & +0.017 \\
    \noalign{\smallskip}\hline
    \end{tabular}%
    }
    \caption{Performance of Trans-Checker on CAIA: BERT's performance is marked as the base numbers $x_p$, $x_r$, $x_f$, and $x_a$. All other numbers are the differences compared to base numbers.}
  \label{tab:errorDetectorAlexa}%
\end{table}%

\subsubsection{Trans-Filler Evaluation}
\label{section_filler}

This ablation study compares four variants of Trans-Filler on the CAIA dataset. The results are shown in Table \ref{tab:ablation_components}. Trans-Filler is fine-tuned on top of BART-base by pairs of annotators and gold transcriptions. Both AR and NAR decoders substantially outperform vanilla BART, and Filler-NAR outperforms the AR setting since autoregressive tends to fill more tokens than needed or generate hallucinations. Further, adding a phoneme embedding to the input embedding layer of BART introduces significant improvements. Overall, Filler+NAR+Pho achieves the best WER, reducing 3.25\% WER compared to annotators. 

\begin{table}[!htbp]
  \centering

     \resizebox{0.48\textwidth}{!}{
    \begin{tabular}{lccccc}
    \hline\noalign{\smallskip}
          & \multicolumn{1}{l}{WER} & \multicolumn{1}{l}{Precision} & \multicolumn{1}{l}{Recall} & \multicolumn{1}{l}{F1} \\
          \noalign{\smallskip}\hline\noalign{\smallskip}
    Annotator & $x_{wer}$ & $x_p$ & $x_r$ & $x_f$ \\
    Baseline BART & +1.06 \% & +0.063 & +0.066 &+0.065\\
    Filler+AR & -1.67\% & +0.166 &+0.172 &+0.169 \\
    Filler+NAR & -2.01\% & +0.165 & +0.173 & +0.169 \\
    Filler+NAR+Pho & \textbf{-3.25\%} & +0.223 & +0.205 &+0.214\\
    \noalign{\smallskip}\hline
    \end{tabular}%
    }
    \caption{Performance of Trans-Filler on CAIA: Precision is the percentage of correctly filled words out of all filled words. Recall is the percentage of correctly filled words out of the number of words that should be filled. Annotator performance is set to base numbers and all other numbers are differences to the bases.}
  \label{tab:ablation_components}%
\end{table}


\subsubsection{The Overall Performance of HTEC}
Table \ref{tab:op} shows the performance of HTEC compared to other methods and human annotators. HTEC uses BERT+Pho as Trans-Checker, and NAR+Pho as Trans-Filler. HTEC achieves the best WER, outperforming human annotators. In comparison, grammatical error correction (GECToR and LM-Critic), ASR error correction methods (ConstDecoder and Rescorer), and Alpaca work worse than human annotators. Below is a one-shot prompt template for Alpaca. In addition, Table \ref{tab:op} presents the improvement contributed by each component: Trans-Checker and Trans-Filler.

\begin{small}
\begin{verbatim}

Instruction: given one utterances and two versions 
of transcriptions from human and ASR model. 
Each transcription may or may not contain errors. 

For example, 
Human: What is on and a half sticks of buttern
ASR: What is on and a half sticks of butter
The correct human transcription is 
'What is one and a half sticks of butter'

Follow the instruction to correct this human 
transcription:
Human: '...'
ASR: '...'

\end{verbatim}
\end{small}

\begin{table}[!htbp]
  \centering
  
  \resizebox{0.45\textwidth}{!}{
  \begin{threeparttable}
    \begin{tabular}{lcc}
    \hline\noalign{\smallskip}
         Method & CAIA WER & MASSIVE WER \\
          \noalign{\smallskip}\hline\noalign{\smallskip}
          Annotator    & $x$ (-) & 17.42\% (-) \\
          ASR & (+24.33\%) & - \\
          Gector & (+34.94\%) & 21.92\% (+25.83\%)\\
          Lm-Critic & (+25.02\%) & 22.89\% (+31.40\%)\\
          BART & (+9.15\%) &  19.07 (+9.50\%) \\
          ConstDecoder & (+2.57\%) & -  \\
          Rescorer & (+2.62\%) & - \\
          Alpaca one-shot & (+6.72\%) & 18.51\% (+6.26\%) \\
          Alpaca few-shot~\tnote{*} & - & - \\
          Trans-Checker & (-1.30\%) & 16.96\% (-2.62\%) \\
          Trans-Filler & (-0.96\%) & 17.08\% (-1.91\%) \\
          HTEC & \textbf{(-2.24\%)} & 16.63\% (\textbf{-4.54\%}) \\
          \noalign{\smallskip}\hline
    \end{tabular}%
    \begin{tablenotes}\footnotesize
    \item[*] Few-shot prompt is presented in Appendix~\ref{app: few-shot}. Alpaca is able to follow simple instruct and one-shot example, but unable to follow complicated instruct such as this few-shot prompt~\ref{app: few-shot}
    \end{tablenotes}
    \end{threeparttable}

    }
    \caption{Overall Performance Evaluation on CAIA and MASSIVE datasets: Annotator performance is set to the base number $x$ for CAIA and all other numbers are average relative numbers compared to the base. ConstDecoder and Rescorer are not applicable to MASSIVE as it does not contain ASR text.}
  \label{tab:op}%
\end{table}%


\subsection{Simulation Study}

\begin{table}[!htb]
  \centering
  
   \resizebox{0.48\textwidth}{!}{
    \begin{tabular}{lccccccc}
    \hline\noalign{\smallskip}
    Annotator & 0 & 10\% & 30\%  & 60\% & 100\% \\
    \noalign{\smallskip}\hline\noalign{\smallskip}
     $x$ & \textbf{(-2.24\%)} & (-4.40\%) & \textbf{(-9.40\%)} & (-18.55\%) & (-28.04\%) \\
    \noalign{\smallskip}\hline
    \end{tabular}%
    }
    \caption{HTEC Simulation with Human-in-the-Loop: Annotators' WER is set as baseline and all others are relative numbers compared to annotator performance.}
  \label{tab:cascadeprecisionhumanintheloop}%
\end{table}%

By assuming Mask Correction Rate (MCR) by annotators, we can dry run a simulation to estimate WER improvement by deploying HTEC to assist annotators during transcription. MCR is the probability that annotators reject false positive error masks or add true positive masks. The higher MRC denotes more expertise of the annotator. Table \ref{tab:cascadeprecisionhumanintheloop} shows HTEC can achieve a significant WER improvement ($9.4\%$) with only $30\%$ MCR assumption. In the next section, we will show the real impact of HTEC as a transcription co-pilot to assist human annotators. Based on the result, we found the MCR is 50\%.


\section{Human-HTEC Collaborative Workflow}
\label{sec:application}
\subsection{User Study: Assist Professional Annotator}

\begin{figure*}[!htbp]
\centering
\includegraphics[angle=0,width=0.99\textwidth,keepaspectratio]{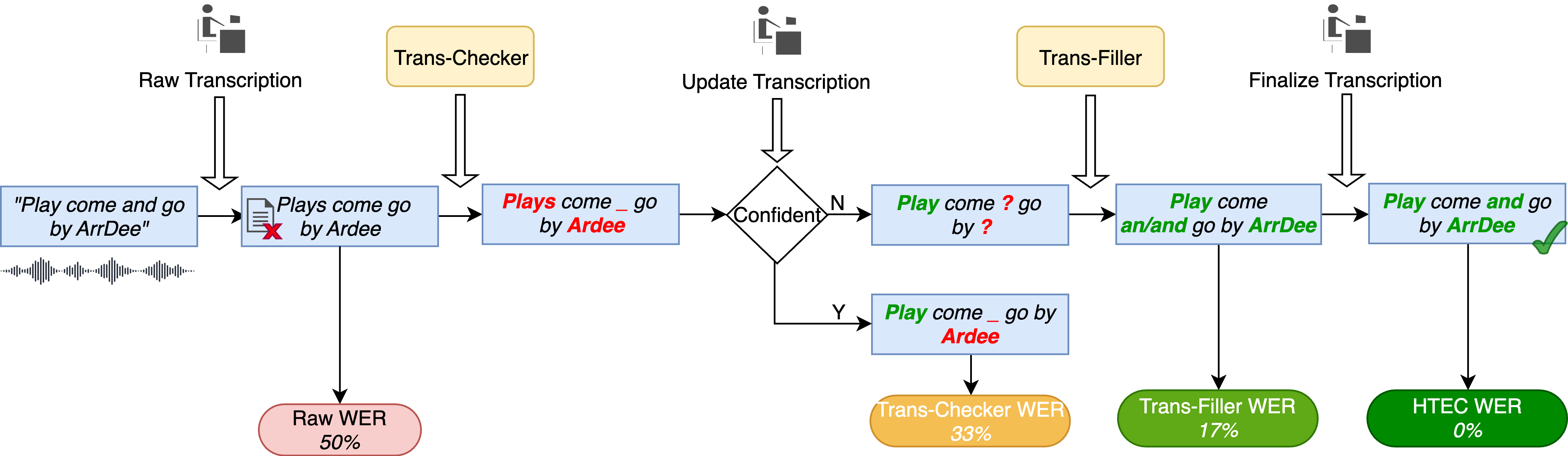}
\caption{Apply HTEC as a co-pilot: Annotator transcribed \textit{``Plays come go by Ardee''}. Trans-Checker highlighted 3 errors, and the annotator was able to fix \textit{``Plays''}. Then the annotator replaced the words that they were not confident enough to transcribe with question marks, which invokes Trans-Filler. The one-best output of Trans-Filler: \textit{``Play come an go by ArrDee''}, corrected one more error: \textit{``Ardee''}. Next, the annotator double-checked the suggestions from Trans-Filler, and picked the correct word \textit{``and''}. Finally, HTEC transcription is correct.}
\label{fig:user}       
\end{figure*}

We conducted a user study to evaluate the impact of HTEC in real-world human transcription. Figure~\ref{fig:user} illustrates the workflow. Annotators first listened to the audio and typed transcriptions. Trans-Checker detected word errors and popped up suggestions that the annotator could take or ignore. Annotators can replace uncertain words with question marks. This would invoke Trans-Filler to fill in a word that the annotator can accept or even make more edits. Annotators could skip Trans-Filler if they did not need the assistance for the audio. For each utterance, we recorded four transcriptions: raw transcription without HTEC's help, updated transcription with the help of Trans-Checker, Trans-filler's output, and final transcription from Trans-filler's output and double-checked by the annotator. Each transcription was compared with the gold transcription that had been previously obtained and was independent of the experiment. Each gold transcription was obtained by the majority vote of three blind-pass annotators. Adjudicators decided ties.

Five professional annotators in US transcribed 1000 utterances. Audios have a length of 2.88 seconds and contain 5.64 words on average. The average voice energy is 49 dB and noise energy averages 15 dB (\ie moderate voice signal with noise). They had 4452 words and 1195 unique words. The annotators have no access to gold transcriptions. The results are presented in Table~\ref{tab:user_study}. Trans-Checker reduces WER by 8.86\%, and Trans-Filler further improves WER by 6.83\%. In total, WER is improved by 15.08\%. We observe that HTEC helps reduce WER for every annotator (Appendix~\ref{app: user_detail}), with a larger improvement for less experienced annotators. In addition, Trans-Filler alone is able to improve WER by 5.36\%.

\begin{table*}[!htbp]
\centering

\resizebox{0.85\textwidth}{!}{
\begin{tabular}{lcccccc}
\hline\noalign{\smallskip}

Error Type & Base WER & Share & Error Prevalence & Trans-Checker & Trans-Filler & HTEC \\
\hline\noalign{\smallskip}

Convention Error & $x_c$ & 5.40\% & 8.57\% & (-17.81\%)  & (-1.67\%)  & \textbf{(-19.18\%)} \\
\hline\noalign{\smallskip}

Spelling Mistake & $x_s$ & 7.33\% & 11.63\% & (-13.98\%)  & (-6.25\%)  & \textbf{(-19.35\%)} \\
\hline\noalign{\smallskip}

Grammatical Error & $x_g$ & 6.94\% & 11.02\% & (-12.99\%)  & (-10.34\%)  & \textbf{(-22.01\%)} \\
\hline\noalign{\smallskip}

Lack Domain Knowledge & $x_l$ & 11.57\% & 18.37\% & (-5.29\%)  & (-1.68\%)  & \textbf{(-6.88\%)} \\
\hline\noalign{\smallskip}

Misheard Audio & $x_m$ & 31.75\% & 50.41\% & (-9.36\%)  & (-9.86\%)  & \textbf{(-18.30\%)} \\
\hline\noalign{\smallskip}

No Error & $x_n$ & 37.02\% & / & (+1.27\%)  & (+0.95\%)  & \textbf{(+2.23\%)} \\
\hline\noalign{\smallskip}

Overall & $x_a$ & 100\% & 100\% & (-8.86\%)  & (-6.83\%)  & \textbf{(-15.08\%)} \\
\hline\noalign{\smallskip}

\end{tabular}
}
\caption{The WERs with Different Levels of HTEC's Assistant: Annotator's performance is marked as a base number. All other numbers are relative to the base WER.}
\label{tab:user_study}
\end{table*}

\subsection{Impact of HTEC to Fairness}
\label{sec:fair}
ASR system often perform insufficiently in tailed and low-resource class~\cite{winata2020adaptandadjust}. Therefore, it is important not to degrade the transcription quality in minor classes, such as utterances from non-native speakers or from rare domains. We analyzed the impact of HTEC by cohort. Results are presented in Table~\ref{tab:fairness}. HTEC improves WER across all cohorts, regardless of the prevalence, domain, locale, or nativity of speakers. WER in minor classes is improved, such as utterances by non-native speakers, shopping domains (e.g., purchase item, query product details), and tailed utterances.

\begin{table}[!htb]
  \centering
  
   \resizebox{0.48\textwidth}{!}{
    \begin{tabular}{llccccccc}
    \hline\noalign{\smallskip}
 &  & Total Words & HTEC WER ($\Delta$) \\
\hline\noalign{\smallskip}
\multirow{3}{*}{Frequency} & Head & 2,129 & (-19.62\%) \\
 & Torso & 1,350 &  (-12.70\%) \\
 & Tail & 15848 &  (-5.46\%) \\
 \hline\noalign{\smallskip}
\multirow{7}{*}{Domain} 
 & Communication & 302 & (-15.16\%) \\
 & Home Automation & 1329 & (-6.80\%) \\
 & Knowledge & 1721 & (-4.50\%) \\
 & Music & 4026 & (-12.51\%) \\
 & Shopping & 935 & (-7.36\%) \\
 & System & 1212 & (-4.42\%) \\
 & Video & 2679 & (-11.15\%) \\
 \hline\noalign{\smallskip}
\multirow{5}{*}{Locale} 
 & Australia & 4,989 & (-8.70\%) \\
 & Canada & 2,428 & (-10.28\%) \\
 & India & 4105 &  (-6.27\%) \\
 & UK & 3,919 &  (-9.66\%) \\
 & USA & 3,886 & (-3.21\%) \\
 \hline\noalign{\smallskip}
\multirow{2}{*}{Nativity} & Native Speaker & 14765 &  (-8.37\%) \\
 & Non-Native Speaker & 4562 & (-5.82\%) \\
 \noalign{\smallskip}\hline
\end{tabular}
}
\caption{Compare WER of HTEC and Annotators' Raw Transcriptions by Cohort of Utterances.}
\label{tab:fairness}%
\end{table}

\section{Conclusion}
\label{sec:conclusion}
This paper studies the problem of improving human transcription quality. We presented the type and cause of the error and found that existing methods are not effective for this problem. We propose HTEC, a two-stage framework for transcription error correction. In the first stage, Trans-Checker detects word errors. In the second stage, Trans-Filler fills in the positions that Trans-Checker detected as errors or that the annotator was not confident enough to transcribe. We further propose a variant of the embedding layer and four novel and simple word editing operations. Experiments show that HTEC improves WER by 2.2\%-4.5\% over human annotators and outperforms other methods by a large margin. We further deployed HTEC as a co-pilot tool to help the annotator in the real-world application. Human-TEC collaborative transcription reduces WER by a relative 15.1\%. For future work, we will improve HTEC by leveraging more acoustic signals in audio-text alignment detection. Another direction is to further explore larger LLM including prompt bootstrapping and more in-context learning experiments.


\clearpage
\section*{Limitations}
\label{sec:Limitations}
HTEC has not been evaluated with non-English data. The performance in other languages and multilingual transcriptions is unclear. Nevertheless, HTEC is applicable to non-English data.

Bias is a prevalent concern with co-pilot tools. HTEC has no exception. We have demonstrated HTEC can lower WER across all cohorts presented in section~\ref{sec:fair}. However, we see that the relative improvement is not uniformly distributed. For example, transcriptions of native speakers' utterances improve more than those of foreign speakers. HTEC is more advantageous for frequent spoken (Head) utterances than for less frequent spoken (Torso and Tail) utterances. These unequal impacts can be balanced by improving training data or applying sampling methods.

Another limitation is that this study has not fully explored LLM as a solution for correcting errors in human transcription. We have presented the evaluation results of Alpaca which performs 10\% worse than HTEC and 6\% worse than annotator. However, we have not compared with larger LLMs, such as PaLM-2~\cite{anil2023palm} or GPT-4~\cite{openai2023gpt4} that are known to be supreme in many tasks. Also, since these larger LLMs can well follow instructions, they may gain further improvement through few-shot learning and careful prompt engineering. Further, exploring the capabilities of LLM is one future direction. We hope to receive feedback and further improve the method, which will help improve human transcription quality.

\section*{Ethics Statement}
\label{sec:Ethics}
\paragraph{Broader Impact} HTEC will be influential in two aspects. Firstly, HTEC will improve human transcription quality. Since the performance of ASR models has been saturated recently on benchmarking datasets but much worse for real-world speech, high-quality transcriptions will help further improve speech recognition models. Secondly, HTEC will largely alleviate the annotator's burden and benefit their mental health. Speech transcription work is challenging, under high pressure, tedious, and potentially harmful to human cognition. Human-HTEC collaborative transcription can ease annotators' work. In addition, we also hope this study can inspire the community to explore more in this direction. 

\paragraph{Ethical Concern} Annotators may over rely on the predictions and recommendations from HTEC. It would be more helpful to deploy a sanity checker and abuse detector during the use of HTEC. Besides, when Human-HTEC generates incorrect transcription that causes damage to the customer, it would be unclear to what extent HTEC should take responsibility.

\paragraph{Human Annotation} All human annotations and transcriptions were conducted by a reputable data annotation provider. The annotators are fairly compensated based on the market price. We did not directly contact the annotators, and we do not have their personal information. They received full training on the annotation policy, and they are aware of the potential risks, such as exposure to harmful content in audio or text data. They are aware of and consent to the use of data.

\paragraph{User Data} We de-identified dataset to protect user privacy. All examples presented are from public datasets or fabricated examples. They are only used for illustration and should not be linked to any demographic or identical information. 

\paragraph{Computational Cost} We use PyTorch in this study. Training HTEC takes 8 GPU hours (Tesla T4). Hosting HTEC as co-pilot requires GPU instances and one instance of Tesla T4 can easily handle 10 requests per second. Hosting HTEC as an autocorrection tool can be asynchronous and CPU instances can well support it. The costs of both the training and hosting HTEC are much cheaper than LLM. Compared to Alpaca, the training cost of HTEC is around 20X cheaper, and the real-time inference cost is approximately 8X cheaper.

\clearpage

\appendix

\section{Appendix}
\label{sec:appendix}

\subsection{Trans-Checker Detail}
\label{app: weight_checker}
While auto error correction requires high precision to mitigate error accumulation, the application with human-in-the-loop desires high recall to identify more errors in the first stage, \ie Trans-Checker. Table~\ref{tab:weight_checker} shows model performance with weighted loss that emphasizes error classes. This inverse-frequency class weighting strategy is beneficial to scenarios with human-in-the-loop to present more true errors to human annotators prior to the stage of Trans-Filler. In contrast, uniform weight, as shown in table~\ref{tab:errorDetectorAlexa} is suitable for autocorrection where only a small proportion of errors of high confidence should be autocorrected.

\begin{table}[!htbp]
  \centering  
  \resizebox{0.48\textwidth}{!}{
    \begin{tabular}{lrrrr}
    \hline\noalign{\smallskip}
          & Precision &  Recall &   F1 &  AUC \\
          \noalign{\smallskip}\hline\noalign{\smallskip}

    BERT-Weighted  & yp & yr  &yf & ya \\
    BART-Weighted  & +0.013 & -0.057  &+0.002& -0.014 \\
    BERT-Pho-Weighted & -0.010 & +0.022  &-0.007 & +0.002 \\
    ELECTRA-Weighted & +0.004 & -0.001 &+0.004 & +0.002 \\
    
    \noalign{\smallskip}\hline
    \end{tabular}%
    }
    \caption{Performance of Trans-Checker with Inverse-frequency Class Weight.}
  \label{tab:weight_checker}%
\end{table}%

\subsection{Data Augmentation}
\label{app: data_augmentation}
The advantage of Filler+AR is that it can handle one-to-many mask filling tasks. Hence, we additionally mask two-gram combinations at random positions of gold transcriptions in the training set to create augmented transcriptions, 5X size of the original dataset, as supplemental training data. We experiment with various ratios of synthetic data mixed with original train data. The synthetic data improves AR but still performs worse than the NAR setting. Notably, the augmentation is not applied to the NAR setting. The reported performance of NAR models in table~\ref{tab:ablation_components} is without data augmentation. A more careful design of synthetic error generation may improve Trans-Filler.

\subsection{Baseline Model Detail and Model License}
GECToR and LM-Critic were originally designed to correct grammatical and spelling errors in written language produced by humans; ConstDecoder and Rescorer are designed to correct errors in ASR hypotheses produced by models. BART is a generic, pre-trained sequence-to-sequence model.

The decoders of GECToR, LM-Critic, ConstDecoder, and BART generate corrected transcriptions directly. The decoder of Rescorer instead renders an ordinal number that indicates which ASR text or annotator transcription should be selected as the corrected transcription. When annotator transcription is selected by Rescorer, it implies the transcription is correct and no correction is needed.

All baseline models used in this study are publicly available and good for research use. Our main models are related to models and tools including BERT, BART, Phoneme embedding model, ERRANT, etc. They are all publicly available and good to use.

\subsection{Trans-Checker Metrics}
We evaluate the precision and recall of Trans-Checker. For example, given that the utterance \textit{``give me the latest news''} was mistakenly transcribed as \textit{``give my latest muse''}, 3 word errors \textit{(``my'', ``the'', ``muse'')} were made. If Trans-Checker detects 3 word errors \textit{(``my'', ``muse'', ``give'')}, both recall and precision are 67\% since 2 out of 3 positives are predicted as positive and 2 out of 3 predicted positives are true positives. Using the same example to illustrate the WER of Trans-Filler, the input to Trans-Filler would be \textit{``give <mask> <mask> latest <mask>''}. If Trans-Filler filled it as \textit{``give me some latest news''}, WER has reduced from 60\% to 20\% as errors \textit{(``my'', ``muse'')} were fixed. 

\subsection{User Study Results per Annotator}
\label{app: user_detail}
\begin{table*}[!htbp]
\centering

\resizebox{0.55\textwidth}{!}{
\begin{tabular}{ccccc}
\hline\noalign{\smallskip}
& Annotator & Trans-Checker & Trans-Filler & HTEC \\
\hline\noalign{\smallskip}
1 & $x_1$ & (-8.80\%) & (-17.52\%) & (-26.32\%) \\
2 & $x_2$ & (-13.84\%) & (-6.36\%) & (-20.19\%) \\
3 & $x_3$ & (-8.92\%) & (-3.46\%) & (-12.37\%) \\
4 & $x_4$ & (-6.99\%) & (-1.59\%) & (-8.59\%) \\
5 & $x_5$ & (-1.48\%) & (+0.73\%) & (-0.74\%) \\
\hline\noalign{\smallskip}
Overall & $x$ & (-8.86\%)  & (-6.83\%)  & (-15.08\%) \\
\hline\noalign{\smallskip}
\end{tabular}
}
\caption{Transcription WER of Each Annotator with Different Levels of HTEC’s Assistant: Annotator’s raw WER is marked as base numbers. All other numbers are relative to the base WERs.}
\label{tab:user_study_per_annotator}
\end{table*}
In section \ref{sec:application}, we show HTEC assists annotators as a co-pilot tool, which reduces 15.1\% WER on average in transcription. The table \ref{tab:user_study_per_annotator} provides step-by-step WER reduction by applying HTEC for each DA.

\subsection{Alpaca Few-shot Prompt}
\label{app: few-shot}
\begin{small}
\begin{verbatim}
Instruction: Given one utterances audio and two 
versions of transcriptions from human and ASR 
model. Each transcription may or may not 
contain errors.

Here are a few examples: 

1. Human transcription has spelling error: 
 Audio: 'What is one and a half sticks of butter'
 Human: 'What is on and a half sticks of buttern'
 ASR: 'What is on and a half sticks of butter'
 
2. Human transcription has grammatical error: 
 Audio: 'Give me, um... the latest news'
 Human: 'Give my, um... the latest news'
 ASR: 'Give the latest news'

3. Human transcription has convention error. 
 Convention errors include format error such as 
 punctuation error or spoken written format error.
 Audio: 'Order triple A. battery'
 Human: 'Order AAA battery'
 ASR: 'Triple A. battery'

4. Human transcription has incorrect entity 
 because the person lacks domain knowledge.
 Audio: 'List ingredients for Mitchell cocktail'
 Human: 'List ingredients for Michelle cocktail'
 ASR: 'List ingredients for cocktail'

5. Human transcription has incorrect contents 
 due to homophone or poor audio quality.
 Audio: 'When was my last Amazon order'
 Human: 'When does my lost Amazon order'
 ASR: 'When was my lost Amazon order'

Correct human transcription in this example:
Human: 'Plays come go by Ardee'
ASR: 'Play come and go by art'

\end{verbatim}
\end{small}

\medskip

\end{document}